\documentclass[]{jfm}
\usepackage{natbib}
\usepackage{amsmath}
\usepackage[dvips]{graphicx}
\usepackage{psfrag}

\newcommand{\aver}[1]{\left\langle {#1} \right\rangle}
\newcommand{\Upower}{U_\Pi}
\newcommand{\Repower}{Re_\Pi}

\begin{document}

\title[Simulation of duct flows with CPI]{Numerical simulation \\ of turbulent duct flows \\ with  constant power input}

\author[Y. Hasegawa, M. Quadrio \& B. Frohnapfel]{
By 
Y\ls O\ls S\ls U\ls K\ls E\ls\ns H\ls A\ls S\ls E\ls G\ls A\ls W\ls A$^1$, \ls 
M\ls A\ls U\ls R\ls I\ls Z\ls I\ls O\ls \ns Q\ls U\ls A\ls D\ls R\ls I\ls O$^2$
\and
B\ls E\ls T\ls T\ls I\ls N\ls A\ls\ns F\ls R\ls O\ls H\ls N\ls A\ls P\ls F\ls E\ls L$^3$
}
\affiliation{
$^1$ Institute of Industrial Science, The University of Tokyo, 4-6-1 Komaba, Meguro-ku, Tokyo 153-8505, Japan \\
$^2$Dipartimento di Scienze e Tecnologie Aerospaziali del Politecnico di Milano \\ via La Masa 34, 20156 Milano, Italy \\
$^3$ Institute of Fluid Mechanics, Karlsruhe Institute of Technology (KIT) \\ Kaiserstr. 10, 76131 Karlsruhe, Germany
}

\maketitle

\begin{abstract}
The numerical simulation of a flow through a duct requires an externally specified forcing that makes the fluid flow against viscous friction. To this aim, it is customary to enforce a constant value for either the flow rate (CFR) or the pressure gradient (CPG). When comparing a laminar duct flow before and after a geometrical modification that induces a change of the viscous drag, both approaches (CFR and CPG) lead to a change of the power input across the comparison.
Similarly, when carrying out the (DNS and LES) numerical simulation of unsteady turbulent flows, the power input is not constant over time. Carrying out a simulation at constant power input (CPI) is thus a further physically sound option, that becomes particularly appealing in the context of flow control, where a comparison between control-on and control-off conditions has to be made. 

We describe how to carry out a CPI simulation, and start with defining a new power-related Reynolds number, whose velocity scale is the bulk flow that can be attained with a given pumping power in the laminar regime. Under the CPI condition, we derive a relation that is equivalent to the Fukagata--Iwamoto--Kasagi relation valid for CFR (and to its extension valid for CPG), that presents the additional advantage of natively including the required control power. The implementation of the CPI approach is then exemplified in the standard case of a plane turbulent channel flow, and then further applied to a flow control case, 
 where the spanwise-oscillating wall is used for skin friction drag reduction. For this low-Reynolds number flow, using 90\% of the available power for the pumping system and the remaining 10\% for the control system is found to be the optimum share that yields the largest increase of the flow rate above the reference case, where 100\% of the power goes to the pump. 

\end{abstract}

\section{Introduction}
\label{intro}

Fluids flow through a duct upon external actions (pressure differences, forces). For the numerical simulation of a duct flow, one needs to mimic these physical effects by supplementing the governing equations with a forcing that drives the flow through the duct against viscous drag, which in case of a straight duct only arises from friction. In practice, unless the wall moves and produces a shear-driven flow, this forcing is imposed by selecting a constant (in time) value for either the pressure gradient or the flow rate. We define the former the Constant Pressure Gradient (CPG) and the latter the Constant Flow Rate (CFR) approach. The two approaches are trivially identical in the laminar flow regime, where flow rate and pressure gradient are uniquely related.

\begin{figure}
\centering
\includegraphics[width=0.8\textwidth]{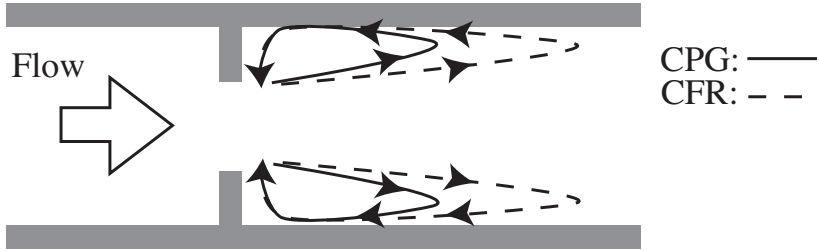}
\caption{Example sketch of the difference between CFR and CPG when comparing the steady laminar flow in a straight reference pipe to the flow in the same pipe with an obstacle. This might be for example the case for studying a stenotic flow in an artery. Depending on whether the mass flow or the pressure gradient of the reference flow is kept constant, the details of the recirculating bubble after the obstacle are different; in particular the recirculating bubble is longer with CFR.}
\label{fig:example}
\end{figure}

Already for steady laminar flows, however, the difference between CFR and CPG becomes visible whenever a comparison between two flows is to be done. If, for example, an obstacle is placed in a duct flow (like an orifice in a hydraulic pipeline system or a stenosis growing in an artery), the changed flow state can be compared to the original one under the assumption that both possess the same flow rate or the same pressure gradient. The corresponding result would of course differ, since the additional flow resistance would lead to an increase of required pressure gradient in the case of CFR or to a decrease of achieved flow rate in the case of CPG. A consequence of this difference is sketched in figure \ref{fig:example}, where it is schematically shown how the length of the recirculation bubble after the obstacle obtained by CFR is larger than that obtained by CPG.

When Direct Numerical Simulation (DNS) of the Navier--Stokes equations, or Large Eddy Simulation (LES), are used to numerically simulate an unsteady turbulent flow even in a simple straight duct, CPG results in a flow rate that fluctuates in time, with a well-defined time average under equilibrium conditions. Similarly, CFR leads to temporal fluctuations of the space-averaged pressure gradient which, after proper time-averaging, becomes constant under equilibrium conditions. Such temporal fluctuations are known to depend on the computational domain size, and become smaller for larger domains \citep{lozanoduran-jimenez-2014}. In any case, CFR and CPG result in the same time-averaged values for pressure gradient and flow rate respectively, provided the integration time is long enough. However, an important difference between the two approaches exists in terms of power input. The power input into the system (which physically might come from the action of a pump) is given by the product of pressure gradient and flow rate. When either pressure gradient or flow rate change in time, the power input is not instantaneously constant. Although this can be regarded as a minor issue in DNS, the instantaneous fluctuations of the power input under CFR and CPG are larger in LES owing to the larger time step size, and might be of some concern if a constant filter width is used in the definition of the subgrid model. Indeed, the fluctuations in the energy flux are reflected in corresponding fluctuations of the filter length scale expressed in viscous (Kolmogorov) units, while the filter width is usually set statically, once and for all, in relation to the mesh size.

\begin{figure}
\psfrag{y}{$y^+$}
\psfrag{u}{$u^+_{rms}$}
\centering
\includegraphics[width=0.9\textwidth]{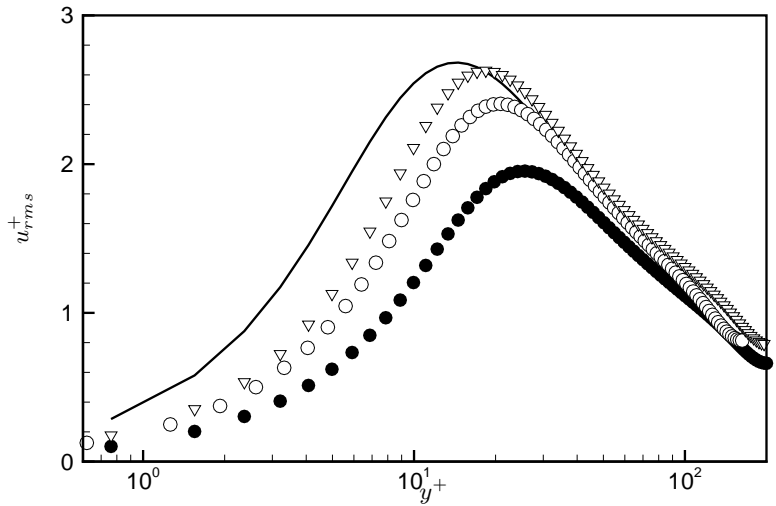}
\caption{Wall-normal distribution of the r.m.s. value of the streamwise velocity fluctuations: effect of the spanwise-oscillating wall. Data from a turbulent channel flow at $Re_\tau=200$ ($Re_b=3173$) taken from \cite{quadrio-ricco-2011}. Continuous line is the reference flow, and symbols correspond to the spanwise-oscillating wall operating with oscillation amplitude $W_0^+=12$ and nearly optimal oscillating period $T^+=100$. Triangles: CPG calculation. Circles: CFR calculation (dataset with filled circles uses friction velocity of the reference flow for non-dimensionalization, while dataset with open circles uses the actual friction velocity in the drag-reduced state).}
\label{fig:interpreting-ow}
\end{figure}

In the context of flow control, where changes in the flow resistance are introduced by design, CFR has been more often employed than CPG,
since evaluating control performance with the resultant drag reduction is more intuitive and the initial transient period to reach an equilibrium state after the onset of control is shorter for CFR than for CPG. From a fundamental viewpoint, the choice between CFR and CPG becomes interesting when the question arises how to
meaningfully compare the controlled flow field to the reference flow without control.
Owing to the action of control, one of the several Reynolds numbers by which the flow can be characterized is modified: using for example a skin-friction drag-reduction technique under CFR has the effect of reducing drag (hence reducing the value of the friction-based Reynolds number, $Re_\tau$, at constant bulk velocity Reynolds number, $Re_b$); however, using the same technique under CPG cannot result in drag reduction, since the drag is constant by definition and the drag-reducing effect manifests itself in an increased flow rate (larger value of $Re_b$ at constant $Re_\tau$). In this case the energy flux (power input) is not constant even in the time-averaged sense: for the controlled flow under CFR the energy flux decreases (pressure gradient decreases and flow rate is unchanged), whereas under CPG the energy flux increases (pressure gradient is unchanged and flow rate increases). It becomes therefore difficult to make any definite statement about the behavior of energy dissipation, i.e. the mechanism by which we lose energy, since the sign of the global change of dissipation is simply {\em prescribed} by the simulation strategy  \citep{ricco-etal-2012}.
In the end, the physical interpretation of results can be quite different depending on which numerical condition is employed to run the simulation.
One example is given in figure \ref{fig:interpreting-ow}, that shows how the drag-reduction technique known as the spanwise-oscillating wall \citep{jung-mangiavacchi-akhavan-1992} affects the r.m.s. intensity of the streamwise velocity fluctuations in a turbulent channel flow. It can be seen that, depending on whether the simulation is run under CPG or CFR (and, in the latter case, depending on whether the friction velocity of the controlled or uncontrolled flow is used to plot the data in viscous units), the global picture of how the oscillations of the wall affect turbulence near the wall can be significantly different: one might rightfully conclude that turbulence intensity is strongly suppressed by the oscillating wall, or that it is essentially unchanged, with perhaps only the profile being slightly shifted outwards.

\begin{figure}
\centering
\includegraphics[width=\textwidth]{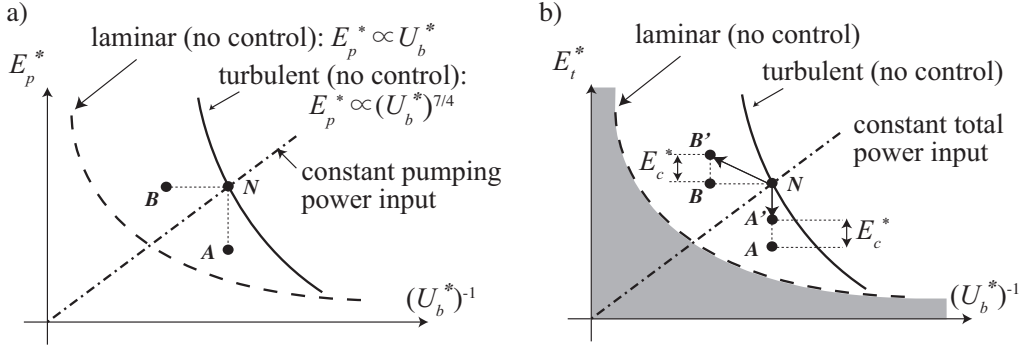}
\caption{Pumping energy, $E_p^*$ (a), and total energy, $E_t^*=E_p^*+E_c^* $ (b), plotted versus the inverse of the dimensional bulk mean velocity, $U_b^*$, for fully developed turbulent flow (solid line) and laminar flow (dashed line). If flow control for skin-friction drag reduction is applied to a turbulent reference point, $N$, the resulting shift in the depicted plane depends on the operating condition of the numerical simulation. The shift to A or A' corresponds to CFR and the one to B or B' to CPG. The vertical distances between A and A' or B and B' in (b) correspond to the energy consumption, $E_c^*$, of the control technique. The dash-dot lines connecting $N$ with the origin of the coordinate system represents states of constant pumping power input, CpPI, in (a) and constant total power input, CtPI, in (b), respectively. Adapted from \cite{frohnapfel-hasegawa-quadrio-2012}.
}
\label{fig:Et_Ub}
\end{figure}

In this paper we propose a strategy to carry out the numerical simulation of duct flow with the power input, i.e. the energy flux through the system, kept constant.
We call this strategy the Constant Power Input (CPI) approach, an alternative to CFR and CPG.
This alternative strategy naturally arises from the money-versus-time plane, a methodology for assessing flow control techniques for skin-friction drag reduction, which we have recently proposed \citep{frohnapfel-hasegawa-quadrio-2012}.
In the money-versus-time plane, the drag reduction problem is generally formulated as finding a compromise between energy consumption and the time required to transport a given amount of fluid over a unit distance, i.e. the inverse of the flow rate. These two essential quantities are plotted on the vertical and horizontal axes of the plane as shown in Fig.~\ref{fig:Et_Ub}. The goal of flow control is to modify a flow state towards the origin of the plane (less energy consumption and more convenience) by manipulating turbulence. By plotting first in Fig.~\ref{fig:Et_Ub}a the pumping energy versus the inverse of the bulk velocity, an uncontrolled fully turbulent flow state is shown to be shifted by successful control into two different end points (both somehow nearer to the laminar state) under CFR and CPG.
Note that the pumping energy per unit fluid volume is proportional to the pressure gradient \citep{frohnapfel-hasegawa-quadrio-2012}, and consequently the flow states under CPG moves along a horizontal line as shown in Fig.~\ref{fig:Et_Ub}a. Since in this plane any straight line starting from the origin connects points with the same pumping power input, control under CPI would therefore move from the uncontrolled flow state along a straight line towards the origin into a third, different final state. In a next step, sketched in in Fig.~\ref{fig:Et_Ub}b, the pumping energy is replaced by the sum of pumping and control energy on the vertical axis, in order to account for the energetic cost of active flow control techniques. Again, it is shown how flow control applied under CFR and CPG shifts the operating point. The straight line between a turbulent reference point and the origin now connects all points with constant total power input. In order to account for the difference of these two planes represented in figure~\ref{fig:Et_Ub} we distinguish between constant pumping power input (CpPI) and constant total power input (CtPI) in the following.
The latter (CtPI) is a specific approach for the evaluation of active flow control techniques; it can be further generalized to account for costs that are related to introducing changes into the fluid system, like geometry or roughness changes.
The former approach (CpPI) does not consider active flow control. It is an alternative simulation strategy to the conventional CFR and CPG approaches, which might be considered more adequate for specific problem formulations.
For example, a flow state realized with a certain pump is described by a characteristic curve
which depends on the principles of the pump employed and also its control scheme. Therefore, the possible trace of the flow state satisfies neither CFR, CPG nor CPI in general. The characteristic curves of most existing turbopumps typically show a decreasing pressure head with increasing flow rate, while the change in the shaft power is rather mild in a wide range of flow conditions. For such a case, CPI might be considered the most reasonable numerical representation.

Our paper begins by first showing in \S\ref{sec:CPI-concept} how the CPI approach can be introduced by identifying a velocity scale, and thus a related Reynolds number, based on the power consumption. The CPI approach is then discussed in the context of the so called FIK identity \citep{fukagata-iwamoto-kasagi-2002} in \S\ref{sec:FIK}, where it is shown how an insightful exact relationship between wall friction (or flow rate) and the turbulent shear stresses can be obtained under CFR, CPG and CPI, and how the latter approach allows the energetic cost of the control to be naturally considered. Next, in \S\ref{sec:CpPI} it is shown how to carry out a numerical simulation under CpPI, whose implementation in a computer code is briefly summarized. Last, the implementation of the CtPI concept is shown in \S\ref{sec:CtPI} to be able to identify, for the simple example of the spanwise-oscillating wall, which percentage of the available power should  be used to run the control in order to achieve the maximum flow rate.

\begin{figure}
\centering
\includegraphics[width=0.7\textwidth]{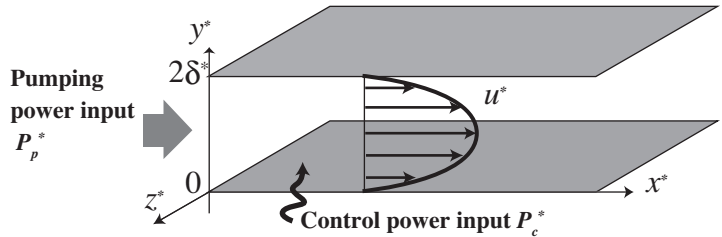}
\caption{Sketch of the plane channel flow geometry. The two walls are located at $y^*=0$ and $y^*=2\delta^*$. The sketch shows the power $P_p^*$ that enters the system through pumping and the power $P_c^*$ required by a flow control device, if present.}
\label{fig:sketch}
\end{figure}

In this paper, all dimensional quantities are indicated with an asterisk. Non-dimension-alization is always based on the velocity scale used in the definition of the relevant Reynolds number. In some instances, viscous (inner) scaling is explicitly denoted with the conventional '+' superscript. The geometrically simple setup of plane channel flow that will be considered in the following is sketched in figure \ref{fig:sketch}; the streamwise, wall-normal and spanwise directions are indicated with $x$, $y$ and $z$, with $u$, $v$ and $w$ being the respective velocity components. The static pressure is $p$. Instantaneous wall-parallel averages of fluid quantities are denoted, for example, by $[u]$, $[p]$ and so forth, which are in general a function of $y$. The volume average is denoted by $\{ \cdot \}$, so that $\{ u \}$ corresponds to the instantaneous bulk velocity. The time average is applied to statistically steady flows. The combination of spatial and temporal averaging  is indicated by angular brackets: $\aver{u}$.

\section{The concept of Constant Power Input (CPI)}
\label{sec:CPI-concept}

We consider one of the simplest canonical flows, namely the fully developed plane channel flow, sketched in figure \ref{fig:sketch}, bounded by two indefinite parallel walls, located $2 \delta^*$ apart. The fluid has density $\rho^*$, dynamic viscosity $\mu^*$ and kinematic viscosity $\nu^*$. In the laminar regime, a given pumping power $P_p^*$ produces a parabolic streamwise velocity profile with a well-defined corresponding mass flow rate. The pumping power per unit wetted area is given by
\begin{equation}
P_p^* = - \frac{d p^*}{d x^*} \delta^* U_b^* ,
\label{eq:Pp}
\end{equation}
where $- d p^* / d x^*$ is the streamwise pressure gradient and $U_b^*$ is the bulk velocity, i.e. the volume-averaged streamwise velocity $\{ u^* \}$. The value of the bulk velocity can be easily deduced by integrating in the wall-normal direction the expression of the laminar parabolic velocity profile:
\begin{equation}
\label{eq:u_bulk_max}
U^*_b = \frac{1}{2\delta^*}\int_0^{2\delta^*} u^*(y^*) \ dy^* =  \frac{1}{3\mu^*} \biggl( -\frac{dp^*}{dx^*}\biggr) {\delta^*}^2.
\end{equation}
In a fully developed flow, the power input to the flow balances the dissipation rate $\epsilon^*$ of the kinetic energy of the flow. The parabolic form of the velocity profile allows $\epsilon^*$ per unit wetted area to be computed as:
\begin{equation}
\label{eq:diss}
\epsilon^* = \int_0^{\delta^*} \mu^* \left(\frac{du^*}{dy^*} \right)^2 dy^* = \frac{3\mu^* {U^*_b}^2}{\delta^*},
\end{equation}
By equating the dissipation rate (\ref{eq:diss}) with the pumping power $P_p^*$, the following relation between the flow rate and the pumping power is found:
\begin{equation}
U^*_b = \sqrt{\frac{P^*_p \delta^*}{3 \mu^*}}.
\label{eq:Ublam}
\end{equation}
This equation identifies a velocity scale, that we indicate with the symbol $\Upower^*$ in the following,  which represents the flow rate (per unit width) achieved by a given pumping power $P_p^*$ in the laminar regime. The flow rate achieved by the same pumping power is of course less in the turbulent regime, where $\Upower^*$ becomes an upper bound. Moreover, theoretical arguments exist \citep{bewley-2009, fukagata-sugiyama-kasagi-2009} showing that $\Upower^*$ is the maximum achievable flow rate for controlled flows if the power consumption of control is  considered.
The velocity scale $\Upower^*$, defined as
\begin{equation}
\Upower^* = \sqrt{\frac{P^*_p \delta^*}{3 \mu^*}}.
\label{eq:Upi}
\end{equation}
is therefore the suitable reference velocity for the CPI approach.

Once the velocity scale $\Upower^*$ is identified, it is straightforward to normalize all quantities by $\Upower^*$ and the channel half width $\delta^*$ in order to obtain governing equations for the CPI concept. In this dimensionless form the incompressible continuity and Navier--Stokes equations read:
\begin{eqnarray}
\frac{\partial u_i}{\partial x_i} &=& 0 \\
\frac{\partial u_i}{\partial t} + \frac{\partial (u_j u_i)}{\partial x_j}
&=& -\frac{\partial p}{\partial x_i} + \frac{1}{\Repower} \frac{\partial^2 x_i}{\partial x_j \partial x_j}
\end{eqnarray}
where the Reynolds number is defined as:
\begin{equation}
\label{eq:Re_p}
\Repower = \frac{\Upower^* \delta^*}{\nu^*}.
\end{equation}
Accordingly, by using (\ref{eq:Upi}) the dimensionless power input is given by:
\begin{equation}
\label{eq:power_normalized}
P_p = \frac{P^*_p}{\rho^*{\Upower^*}^3} = \frac{3}{\Repower}.
\end{equation}
Therefore, $\Repower$ defines the power input into the system and is the natural choice for a CPI simulation, just like $Re_b$ defines the flow rate and is conveniently used in CFR, and $Re_\tau$ defines the pressure gradient and is used in CPG. The relation between the different Reynolds numbers and corresponding velocity scales is summarized in the Appendix.

In the case of active control where the total power input into the fluid system consists of pumping and control power,
\begin{equation}
P_t^* = P_p^* + P_c^*,
\end{equation}
Eq.~(\ref{eq:power_normalized}) becomes
\begin{equation}
P_t = \frac{3}{\Repower}
\label{eq:Pt}
\end{equation}
and states that the numerical value of $\Repower$ sets the total power input. Depending on the particular control, the relative importance of the terms $P_p^*$ and $P_c^*$, i.e. the amount of the available power that is actually used for pumping and the one that goes into the flow control device, is described by the quantity
\begin{equation}
\gamma = \frac{P_c^*}{P_t^*} = 1 - \frac{P_p}{P_t}
\label{eq:gamma}
\end{equation}
which reflects the percentage of the total power that is used for control. Accordingly,
\begin{equation}
\label{eq:Pp_active}
P_p = P_t \left(1 - \gamma\right) = \frac{3\left(1-\gamma\right)}{\Repower}.
\end{equation}

\section{Flow control under CFR, CPG and CPI}
\label{sec:FIK}

A few years ago \cite{fukagata-iwamoto-kasagi-2002} derived the so-called FIK identity, which relates the skin-friction coefficient in a fully developed channel flow to its dynamical contributions under CFR. Since it nicely shows how the wall-normal distribution of the Reynolds shear stress contributes to the wall friction quantitatively, the FIK identity has been widely used for analyzing different drag reduction mechanisms and also for developing novel control strategies.  This idea has later been extended to CPG by \cite{marusic-joseph-mahesh-2007}. After briefly recalling these two results, we present the corresponding formulation for the CPI case for which we also obtain the result that the success of a flow control technique in a turbulent channel flow can be quantified by the achieved reduction of the Reynolds shear stress.

\subsection{CFR}
The original FIK identity is valid under CFR, hence the relevant Reynolds number is $Re_b \equiv U_b^* \delta^*/\nu^*$, and velocities are scaled with the bulk velocity $U_b^*$. It reads:
\begin{equation}
C_f = \frac{6}{Re_b} + 3 \int_0^{2}(1-y) \aver{-u'v'} dy .
\label{eq:FIK_CFR}
\end{equation}
The first term on the r.h.s. corresponds to the laminar drag at the same flow rate, whereas the second term represents the turbulence contribution. Typically, since $\left< -u'v' \right> > 0$, the turbulent shear stress enhances the friction coefficient $C_f$.

\subsection{CPG}
The corresponding relationship for CPG has been derived by \cite{marusic-joseph-mahesh-2007}. The relevant Reynolds number is $Re_\tau \equiv u^*_\tau \delta^*/\nu^*$, and velocities are scaled with friction velocity $u_\tau^*$. It reads:
\begin{equation}
U_b = \frac{Re_\tau}{3} - \frac{Re_\tau}{2} \int_0^2 (1-y) \aver{-u'v'} dy .
\label{eq:FIK_CPG}
\end{equation}
In analogy with Eq.~(\ref{eq:FIK_CFR}), the bulk mean velocity under CPG is composed of two terms, i.e., laminar and turbulence contributions. The first term on the r.h.s. is equal to the laminar flow rate under the same pressure gradient. The second term represents the momentum loss due to the turbulent shear stress leading to a lower flow rate in turbulent flows. This tendency is enhanced at higher Reynolds number.

\subsection{CPI}
\label{sec:FIK-CPI}
Under CPI conditions the relevant Reynolds number is $\Repower \equiv \Upower^* \delta^*/\nu^*$, and velocities are scaled with velocity $\Upower^*$. The averaged streamwise component of the momentum equation can be written as:
\begin{equation}
0 = - \frac{d \aver{u'v'}}{d y} - \aver{ \frac{d p }{d x} } +
\frac{1}{\Repower} \frac{d^2 \aver{u} }{d y^2}.
\end{equation}
Triple integration yields:
\begin{equation}
U_b =  - \aver{ \frac{d p }{d x} } \frac{\Repower}{3}
- \frac{\Repower}{2} \int_0^2 (1-y) \aver{-u'v'} dy.
\label{eq:FIK_CPI}
\end{equation}
This expression for the bulk mean velocity again consists of two terms on the r.h.s.: a laminar contribution and a turbulence contribution. However, in contrast to (\ref{eq:FIK_CFR}) and (\ref{eq:FIK_CPG}) the first term on the r.h.s. is not defined solely by the prescribed Reynolds number but also by the resulting mean pressure gradient term. Hence, it needs to be considered further in order to obtain an expression that only contains the relevant Reynolds number and the Reynolds shear stress.

The mean pressure gradient is given by the pumping power per unit wetted area divided by the bulk velocity; thanks to Eq. (\ref{eq:Pp_active}) it reads
\begin{equation}
-  \aver{ \frac{d p }{d x} } = \frac{P_p}{U_b} = \frac{3(1-\gamma)}{\Repower U_b}.
\label{eq:PG_CPI}
\end{equation}
Notice how $\gamma$ has been introduced into the equation at this stage already, since one of the CPI advantages is to be able to discriminate between pumping power and total power. Using this relationship to replace the mean pressure gradient term in Eq.~(\ref{eq:FIK_CPI}) and multiplying by $U_b $ (for $U_b \ne 0$) results in a quadratic equation for $U_b$:
\begin{equation}
U_b^2 + T U_b -\left(1 -\gamma\right) = 0,
\label{eq:PG_CPI_2}
\end{equation}
where the coefficient $T$ is defined as
\begin{equation}
T = \frac{\Repower}{2} \int_0^2 \left( 1-y \right) \aver{-u'v'} dy
\end{equation}
and represents the contribution of turbulence to the resulting flow rate.
The solution to Eq.(\ref{eq:PG_CPI_2}) is given by
\begin{equation}
U_b = \frac{-T \pm \sqrt{T^2 + 4(1 - \gamma)}}{2}
\label{eq:FIK_CPI_UB}
\end{equation}
and can be regarded as the FIK identity for the CPI condition.

It should be noted that, in the CPI condition, once $0 < \gamma < 1$ is given,
the pumping power is fixed, whereas the flow direction is not prescribed.
The pumping power being, as per Eq.~(\ref{eq:Pp}), the product of pressure gradient and bulk mean velocity explains why two solutions appear in (\ref{eq:FIK_CPI_UB}): one corresponds to the case where both, the bulk mean velocity and the pressure gradient, are positive, and the other to the case where they are both negative. However, choosing the $x$ axis such that the mean flow is oriented in its positive direction, as figure \ref{fig:sketch} suggests, implies that one of the two solutions in Eq.~(\ref{eq:FIK_CPI_UB}) is selected, namely that with the positive sign, to satisfy Eq. (\ref{eq:PG_CPI_2}).

It is evident that for a standard laminar flow, where the power input for control and the turbulence contribution vanish, namely $\gamma = 0$ and $T = 0$, the above equation reduces to $U_b = 1$ or $U_b^* = \Upower^*$, which is the upper limit for the bulk velocity under the CPI condition. When the flow is turbulent, the Reynolds shear stresses are typically negative, i.e. $\aver{-u'v'} > 0$ so that $T >0$, thus leading to $U_b<1$. In the limiting case where the turbulence contribution becomes infinitely large, i.e., $T \rightarrow \infty$, $U_b$ tends to zero. Another limiting behaviour is the case with $\gamma=1$, i.e. all available power goes into the control system. In this case, the quadratic equation~(\ref{eq:PG_CPI_2}) becomes singular, so that
it reduces to the original linear equation~(\ref{eq:FIK_CPI}), which possesses the unique solution $U_b = -T$.
Depending on the particular control strategy, some flow rate may be produced by control through $T \ne 0$, like for example the forcing scheme devised by \cite{min-etal-2006}, where a viscous streaming mechanism provides the equivalent of a pumping action. 

\section{Implementing CPI into a DNS code: CpPI}
\label{sec:CpPI}

We consider first the no-control case, where the power input to the system is entirely given by the pumping power, i.e. $\gamma=0$, and describe how to implement the Constant pumping Power Input (CpPI) concept into a DNS or LES code in the simple geometry of the plane channel flow.

In a CPI simulation the pumping power $P^*_p$ is prescribed, and set by the value of $\Repower$. Compared to a laminar simulation, both the space-averaged instantaneous streamwise pressure gradient $- [\partial p / \partial x]$ and the instantaneous flow rate $\{ u \}$ may fluctuate around their mean values $G$ and $U_b$, but their product sets the instantaneous value of the pumping power and must consequently remain constant to satisfy Eq.~(\ref{eq:power_normalized}). Both, $- [\partial p / \partial x]$ and $\{ u \}$, are thus computed as time-dependent quantities during the simulation.

However, it is not easy to satisfy the CpPI condition exactly at every time step, since
the flow state is obtained as a result of the computation. Hence, we follow the simplest approach in a time-discrete setting, where the equations of motion are advanced by a typically very small computational time step. To advance the solution from time step $n$ to time step $n+1$, the pressure gradient is determined as:
\begin{equation}
- \left[ \frac{\partial p}{\partial x} \right]^{(n+1)} = \frac{P_p}{ \{ u \}^n} = \frac{3}{\Repower \{ u \}^n}
\end{equation}
and this enables advancing the solution by one time step to compute $\{ u \}^{n+1}$.
Note that the above scheme is  first-order accurate in time.
In the present study, it is sufficient, since the estimation error of $\{ u \}^{n+1}$ is commonly in the order of $10^{-8}$.
This indicates that the CPI condition is satisfied within the same order.
The order of accuracy can be increased in a straightforward way by using past variables. For the second-order accuracy, ${u}^{n+1} \approx 2{u}^{n}-{u}^{n-1}$
can be used. We confirmed that the increased accuracy of the numerical scheme does not affect the present results.

When the flow reaches an equilibrium state, the mean pressure gradient and the wall friction must balance:
\begin{equation}
-  \aver{ \frac{d p }{d x} } = G = \frac{1}{\Repower} \left. \frac{d\aver{u}}{dy}\right|_{y=0} .
\end{equation}

This leads to
\begin{equation}
\label{eq:Pt_NC}
P_p  = G U_b =  \frac{ U_b}{\Repower} \left. \frac{d\aver{u}}{dy} \right|_{y=0} .
\end{equation}
Here, it is assumed that the temporal correlation between $G$ and $U_b$ is negligibly small.
This is generally valid in a sufficiently large computational domain except for the special case,
where the pressure gradient is actively varied in time, so that the correlation between
$G$ and $U_b$ becomes significant.

Since $P_p = 3/\Repower$ from Eq.~(\ref{eq:power_normalized}), the following relationship is obtained:
\begin{equation}
U_b \left. \frac{d\aver{u}}{dy} \right|_{y=0} = 3
\label{eq:exact-relation}
\end{equation}
which is exact and can be used to verify that the time averaging is sufficient.

The friction coefficient $C_f$ can be defined in terms of pumping power by:
\begin{equation}
C_f = \frac{P_p^*}{\frac{1}{2}\rho^* U_b^{*3}}
\label{eq:Cf-power}
\end{equation}
and by using Eq. (\ref{eq:Ublam}) it can be expressed as:
\begin{equation}
C_f = \frac{6 \mu^* \Upower^{*2}}{\rho^* \delta^* U_b^{*3}} =
\frac{6}{\Repower} \left( \frac{\Upower^*}{U_b^*} \right)^3.
\label{eq:Cf-nocontrol}
\end{equation}
When the flow is laminar, $U_b^* = \Upower^*$ and thus $Re_b = \Repower$; in this case the above expression reduces to the conventional laminar formula $C_f=6/Re_b$.

\subsection*{Example}
We now present the results of a sample CpPI calculation of (uncontrolled) turbulent channel flow. The statistics are the result of a DNS simulation, carried out by adapting an existing DNS code to the CPI procedure. The code \citep{luchini-quadrio-2006} solves the incompressible Navier--Stokes equations, written in terms of wall-normal velocity and wall-normal vorticity components, by a pseudo-spectral method, where Fourier expansions are employed in the streamwise and spanwise directions, and compact fourth order explicit finite differences schemes are used in the wall-normal direction. Full details can be found in the original paper.
The code changes that we had to implement in order to enable the simulation to run under CPI are extremely limited (five lines of source code).

\begin{figure}
\centering
\psfrag{t}{$t \Upower / \delta$}
\psfrag{Y}{$\{ u \}/U_b, - [\partial p / \partial x]/G$}
\includegraphics[width=0.9\textwidth]{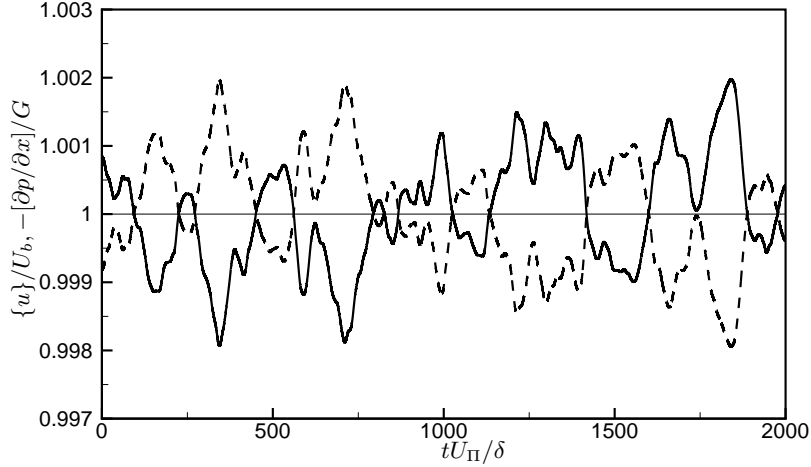}
\caption{Temporal behaviour of the space-mean instantaneous bulk velocity $\{ u \}$ (thick solid line) and the space-mean instantaneous pressure gradient $- [\partial p / \partial x]$ (dashed line), normalized with their time-average values $U_b$ and $G$. The thin horizontal line is the product of the two quantities, proportional to the instantaneous power input, which actually remains constant during the simulation.}
\label{fig:constant-power}
\end{figure}

The  streamwise and spanwise dimensions of the computational domain are set at the standard values of $L_x^* = 4 \pi \delta^*$ and $L_z^* = 2 \pi \delta^*$, as in \cite{kim-moin-moser-1987}. We set the value of the Reynolds number based on the total power input at $\Repower = 6500$, that yields $Re_\tau \approx 200$. The numbers of modes employed in the streamwise, wall-normal and spanwise directions are $(N_x, N_y, N_z) = (256, 128, 256)$, respectively. If expressed in wall units, the spatial resolution of $\Delta x^+= 9.8$, $\Delta z^+=4.9$ and $\Delta y^+= 0.9 - 4.9$ is within the usually accepted values \citep{hoyas-jimenez-2008}. Time integration is carried out via a classic partially implicit 3-stages Crank-Nicholson Runge-Kutta scheme; integration time is $2000 \delta/\Upower$, corresponding to more than 12,000 viscous time units.

Figure \ref{fig:constant-power} demonstrates the temporal behaviour of the flow rate and the pressure gradient. They are normalized with their time-averaged values $U_b$ and $G$.
The figure graphically conveys the message that the product of the two quantities is actually kept constant at every time instant during the simulation. The product of the corresponding bulk velocity and wall shear stress
is 2.99914,  which indicates that Eq.(\ref{eq:exact-relation}) is satisfied to a very high degree.
Additionally, it can be also appreciated how the time scale of variations of both quantities is $O(100)$ and thus far larger than the time step of the simulation, which is of the order of $10^{-2}$.

\begin{figure}
\psfrag{y}{$y^+$}
\psfrag{U}{$\aver{u}/\Upower$}
\psfrag{V}{$\aver{u}^+$}
\psfrag{u}{$u_{i,rms}/\Upower$}
\psfrag{v}{$u_{i,rms}^+$}
\centering
\includegraphics[width=\textwidth]{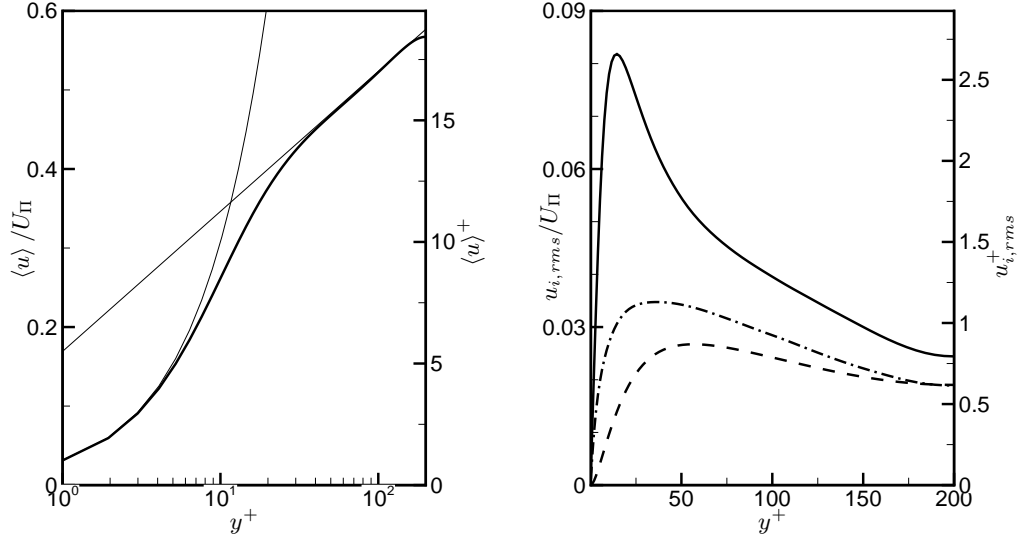}
\caption{Mean velocity profile $\aver{u}$ (left) and wall-normal distribution of the r.m.s. values of velocity fluctuations $u_{i,rms}$ (right) for the DNS of a turbulent channel flow under CpPI. In each plot, the left vertical axis is in "power" units, and the right vertical axis is in the conventional viscous (plus) units. In the left plot, the thin lines are the linear and logarithmic laws given by $\aver{u}^+ = y^+$ and $\aver{u}^+ = 2.5 \ln y^+ + 5.5$. In the right plot, the streamwise component is plotted with continuous line, the wall-normal component with a dashed line, and the spanwise component with a dash-dotted line.
}
\label{fig:sample-results}
\end{figure}

The resultant bulk and friction Reynolds numbers are found to be $Re_b = 3176$ and $Re_\tau = 199.7$, respectively. The obtained mean velocity and r.m.s. values of velocity fluctuations normalized by $\Upower^*$ are shown in Fig.~\ref{fig:sample-results}. The right vertical axis shows the same quantities made dimensionless with the conventional viscous wall units, to highlight the scale factor involved.

\section{Constant total Power Input (CtPI)}
\label{sec:CtPI}

Active flow control techniques require power to operate. As described in \S\ref{sec:CPI-concept}, the additional control power $P_c^*$ is a further contribution to the total power input $P_t^*$ besides the pumping power $P_p^*$.

The procedure previously sketched in \S\ref{sec:CpPI} to adjust flow rate and pressure gradient during the simulation must be adapted as follows. Once the bulk velocity $\{ u \}^n$ and the control power $P_c^n$ are known at the time step $n$, the pumping power at timestep $n+1$ is determined as $P_p^{n+1} = P_t - P_c^n$, and using Eq.~(\ref{eq:Pp_active}), the pressure gradient is determined as
\begin{equation}
- \left[ \frac{\partial p}{\partial x} \right]^{(n+1)}  =
\frac{P_p^{n+1}}{\{ u \}^n} = \frac{P_t - P_c^n}{ \{ u \}^n} =
\frac{3\left( 1 - \gamma \right)}{\Repower \{ u \}^n} .
\end{equation}

Eq.(\ref{eq:Pt_NC}) is still valid (in equilibrium conditions). Hence,
\begin{equation}
P_t \left( 1 - \gamma \right) =
\frac{U_b}{\Repower} \left. \frac{d\aver{u}}{dy} \right|_{y=0}.
\end{equation}

By using (\ref{eq:Pt}) one obtains
\begin{equation}
\left. U_b \frac{d\aver{u}}{dy} \right|_{y=0} = 3 \left( 1 - \gamma \right)
\end{equation}
which is an exact relation, like Eq.(\ref{eq:exact-relation}).

The friction coefficient is still defined in terms of $P_p^*$:
\begin{equation}
C_f = \frac{P_p^*}{\frac{1}{2}\rho^* U_b^{*3}} =
\frac{2 P_t^* ( 1 - \gamma)}{\rho^* U_b^{*3}} = \frac{6}{\Repower} \left( \frac{\Upower^*}{U_b^*} \right)^3 ( 1 - \gamma ) .
\end{equation}

When no control is applied, i.e. $\gamma = 0$, and the flow is laminar, the conventional laminar relationship $C_f = 6/Re_b$ is recovered. \cite{frohnapfel-hasegawa-quadrio-2012} introduced the notion of an effective wall friction $\tau_w^{e *}$ (and an effective friction coefficient $C_f^e$) to account for the total power consumption:
\begin{equation}
\tau_w^{e *} = \frac{P_t^*}{U_b^*} = \tau_w^* + \frac{P_c^*}{U_b^*} .
\end{equation}

By writing it as ${\tau^*_w}^e = \tau^*_w/(1-\gamma)$, we obtain:
\begin{equation}
C_f^e = \frac{C_f}{1 - \gamma} = \frac{6}{\Repower} \biggl(\frac{\Upower^*}{U^*_b}\biggr)^3,
\end{equation}
indicating that achieving $U^*_b = \Upower^*$ is equivalent to reducing $C^e_f$ to the laminar value.

\subsection*{Example}

To exemplify the application of the CPI concept in the context of flow control, we consider the problem of reducing the turbulent friction drag, and employ to this purpose the well-known spanwise-oscillating wall technique \citep{jung-mangiavacchi-akhavan-1992}. The choice of this particular technique is motivated by the amount of information already available, as well as by the limited number of control parameters involved. The forcing scheme consists of harmonically moving the wall in the spanwise direction according to:
\begin{equation}
w(x,y=0,z,t) = W_0 \sin \left( \omega t \right)
\end{equation}
where the only two parameters are $W_0$, the oscillation amplitude, and $\omega = 2 \pi / T$, its frequency ($T$ is the oscillation period). The oscillating wall obviously requires power for its actuation, and thus the question we are asking in the CPI setting is: For a given available total power, which is the optimal share of power between the pump and the control device that maximizes the flow rate? In other words, for a given value of $\Repower$, we search the optimal value of $\gamma$ for which the oscillating-wall technique provides the largest increase of the flow rate $U_b$ above the value $U_{b,0}$ corresponding to $\gamma=0$, the uncontrolled flow where all of the power is used for pumping. It is known \citep{baron-quadrio-1996} that, for relatively small values of the forcing amplitude, the oscillating-wall technique can achieve small positive net savings, i.e. combinations of parameters exist where control power is smaller than pumping power saved thanks to the control action. We thus expect $U_b$ to be maximized by some optimal non-zero value of $\gamma > 0$.

A general optimization procedure would of course involve looping over different values of $\gamma$ in order to identify the best. Moreover, a single value of $\gamma$ corresponds to several different combinations of the control parameters, the optimal set of which has to be identified. An additional difficulty is that, given a set of control parameters, the value of $\gamma$, i.e. the power cost of the control technique in that particular configuration, is in general unknown beforehand, thus calling for an iterative procedure.

However, in the particular case of the oscillating wall, previous knowledge can be exploited, and available information \citep{quadrio-ricco-2004} allow us to determine the range of parameters where best performance in terms of net saved power is expected. Moreover, it is known \citep{quadrio-sibilla-2000} that the oscillating spanwise boundary layer created by the oscillations obeys the laminar solution of the Stokes second problem in the range of interesting forcing conditions: it can thus be anticipated that the control power increases with the square of the oscillation amplitude and with the square root of the oscillation frequency. The values of $\gamma$ corresponding to a given pair of $W_0$ and $\omega$ can thus be first-guessed {\em a priori}, and then simply verified or slightly adjusted {\em a posteriori}.

\begin{figure}
\centering
\psfrag{g}{$\gamma$}
\psfrag{Y}{$ U_b / U_{b,0}$}
\includegraphics[width=0.9\textwidth]{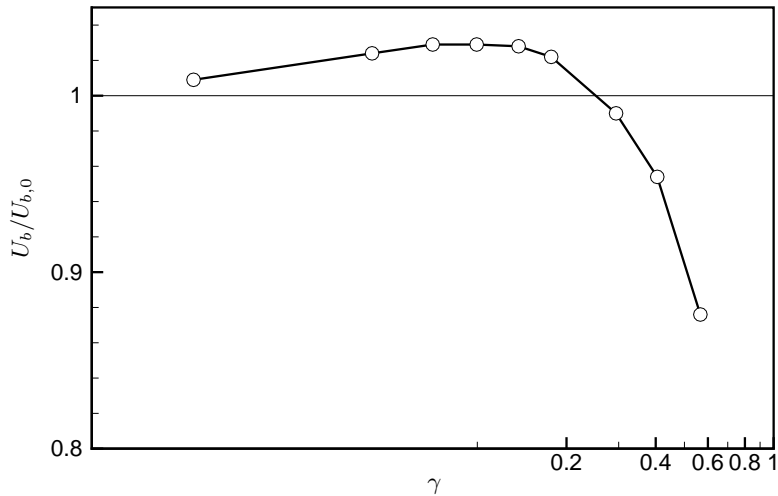}
\caption{Ratio between the actual flow rate $U_b$ and the flow rate $U_{b,0}$ of the uncontrolled case, as a function of the share $\gamma$ of the available power between pumping power and control power provided to the oscillating walls.}
\label{fig:Ub-vs-gamma}
\end{figure}

Figure \ref{fig:Ub-vs-gamma} summarizes the small parametric study carried out for the oscillating wall in the present work. In these simulations, we have chosen the same value $\Repower=6500$ previously used in \S\ref{sec:CpPI}, as it corresponds to the uncontrolled flow at $Re_\tau=200$ where most of the available drag reduction data for the oscillating wall belong. The same computational parameters mentioned before are employed here. In calculating time-average quantities, the initial transient is carefully excluded. 
We note, in passing, that the duration of this transient is one of the outcomes that is most affected by the simulation strategy, with CFR presenting the shortest transient. However, as discussed by \cite{ricco-etal-2012}, the longer transient shown by CPG is roughly compensated by the smaller fluctuation of the quantity to be averaged in time (flow rate in CPG and wall friction in CFR), so that the two approaches are roughly equivalent in terms of computational cost. CPI presents an intermediate behaviour in this respect.

The figure describes how the ratio $U_b / U_{b,0}$ changes with $\gamma$. The oscillating wall is confirmed to possess a parameters range where the optimal share of power between pumping and control power yields an increase in the flow rate.
Best performances are observed near  $\gamma \approx 0.1$, indicating that the optimal use of the available power (for the investigated Reynolds number) is to employ 90\% for the pumping system and the remaining 10\% for the control system. The maximum control-induced increase of the flow rate is rather small, of the order of 3\%, and this is in agreement with previous information \citep{quadrio-ricco-2004} that the maximum net power saving is only 7\%. Of course better-performing drag reduction techniques are expected to yield larger increase of the flow rate.

The calculations at $\gamma>0.25$ do not provide a positive yield for the oscillating wall. When such a control input is applied in CFR or CPG simulation, the resultant drag reduction rate (CFR) or flow rate (CPG) is enhanced. However,
the control power input increases more rapidly. This important point is directly visible for the CPI condition, where the fixed total power input results in a decrease of the flow rate.

In the limit for $\gamma \rightarrow 1$, all the energy is used up by the control system, and in the present case the flow rate decreases to zero. As mentioned in \S\ref{sec:FIK-CPI}, however, this is not necessarily true in general.

\section{Conclusion}
\label{conclusion}

An original approach to carry out numerical simulations of duct flows, that distinguishes itself from the well established options of keeping constant the flow rate (CFR) or the pressure gradient (CPG), is introduced. The approach, that we name CPI, consists of keeping the power input to the system strictly constant in time, thus ensuring a constant energy flux. CPI represents a third alternative to the well established CFR and CPG techniques.

The paper describes with some detail how to implement the CPI concept into a DNS or LES computer code, describing the simple case of a plane channel flow. The idea is general though, as it is based on the identification of a velocity scale $\Upower$ that expresses the flow rate achievable in the laminar regime with a given power input, and leads to the definition of a new power-based type of Reynolds number $\Repower$, which is naturally suited for CPI calculations, much as $Re_b$ is flow-rate based and suited for CFR, and $Re_\tau$ is wall-friction based and suited for CPG.

The application example of CPI given in the paper concerns the problem of turbulent skin-friction drag reduction through the spanwise-oscillating wall. Drag reduction is often motivated by the need to save power. The CPI concept naturally leads to setting up a power-constrained problem where the optimal management of the power budget is identified by finding the optimal share of power between the pump and the flow control device providing the largest flow rate.
A plot of the gain (the flow rate) against the share of the available power between pumping and control provides a simple and effective tool to assess different drag reduction techniques and thus determine which ones should be studied and developed further.

We believe that the potential of the CPI concept goes beyond the applications presented in the paper. For turbulent drag reduction, the CPI concept has the potential to highlight general properties of flows with drag reduction.
It might be expected that drag-reduced turbulent flows with different control strategies share some peculiar and yet universal feature, which must possess a statistical footprint.
However, identifying such a footprint requires comparing flows with and without drag reduction, and the terms of such comparison must be clearly defined. We believe that the CPI concept is a key property of this definition. In the Richardson--Kolmogorov theory of homogeneous isotropic turbulence it is well known that what happens in the inertial range of scales is determined, at statistical equilibrium, by the rate at which energy enters the system.
This corresponds to the rate at which, on average, energy is transferred to smaller scales, and to the mean dissipation rate of turbulent kinetic energy that takes place at the Kolmogorov scale. In the more complicated setting of inhomogeneous, anisotropic turbulence, the CPI approach will allow us to analyze and compare controlled flows where the rate at which energy enters the system is fixed, thus allowing a meaningful analysis of how energy distributes in the system.

From a more general standpoint, the CPI concept could play an interesting role whenever a comparison between two (laminar or turbulent) flows where a modification is introduced has to be carried out. If added to the comparison of two flow fields shown in figure \ref{fig:example}, CPI represents a third alternative (where e.g. the size of the resulting recirculation bubble is in between the two for CPG and CFR) whose meaning must be evaluated on the basis of the specific problem. Any extension of the presented CPI concept to unsteady flow conditions, e.g. in analogy to varying flow rates in arteries, can be considered.

As a final note, we would like to state that whenever fundamental studies on laminar to turbulent transition are concerned it can be envisaged that the way simulations are run (CFR, CPG or CPI) directly influences the result. CPI might therefore be an interesting option to open up the parameter space in this context.

\section*{Acknowledgments}
The authors would like to thank Anna Slotosch for providing the idea to figure 1, and Marco Carini for useful discussions. Y.H. acknowledges the support by the Ministry of Education, Culture, Sports, Science and Technology of Japan (MEXT) through the Grant-in-Aid for Scientific Research (B) (No.25289037). B.F. acknowledges the support of project FR2823/2-1 of the German Research Foundation (DFG). 

\bibliographystyle{jfm}

\appendix
\section{CPI vs CFR vs CPG}

The relation between the newly introduced $\Repower$ for CPI and the existing Reynolds numbers customarily employed for a turbulent channel flow, i.e. the friction and the bulk Reynolds number, is discussed in the following.

For a fully developed turbulent channel flow, the friction coefficient can be estimated by the so-called Dean's formula \citep{dean-1978} as:
\begin{equation}
\label{eq:Cf_Reb}
C_f \equiv \frac{\tau^*_w}{(1/2)\rho^* {U^*_b}^2} = \alpha Re_b^{-1/4} ,
\end{equation}
where the coefficient $\alpha$ takes the value of $\alpha=0.0614$.

Based on this well-established empirical correlation, the relationship between $Re_\tau$ and $Re_b$ is given by:
\begin{equation}
\label{eq:Ret_Reb}
Re_\tau = \sqrt{\frac{\alpha}{2}} Re_b^{7/8} = 0.1752 Re_b^{7/8}.
\end{equation}
After (\ref{eq:Cf-power}), the pumping power $P^*_p$ is related to $C_f$ by:
\begin{equation}
\label{eq:P_p_1}
P^*_p = \frac{1}{2}\rho^* {U^*}^3_bC_f.
\end{equation}
Substituting $P^*_p$ by Eq.~(\ref{eq:power_normalized}) yields:
\begin{equation}
\label{eq:power_turb}
\frac{\Upower^*}{U^*_b} = \frac{\Repower}{Re_b} = \sqrt{ \frac{1}{6} Re_b C_f} = 0.1012 Re_b^{3/8},
\end{equation}
and therefore
\begin{equation}
\label{eq:Rep_Reb}
\Repower = 0.1012 Re_b^{11/8}.
\end{equation}
Eq.~(\ref{eq:P_p_1}) can also be written as:
\begin{equation}
\label{eq:P_p_3}
P^*_p = \rho^* {u^*_{\tau}}^2 U^*_b.
\end{equation}
Combined with Eqs.~(\ref{eq:P_p_3}) and (\ref{eq:power_normalized}), the following relationship is obtained:
\begin{equation}
\frac{\Upower^*}{u^*_\tau} = \frac{\Repower}{Re_\tau} = \sqrt{\frac{Re_b}{3}}
= 1.5621 Re_\tau^{4/7}.
\end{equation}
Hence,
\begin{equation}
\Repower = 1.5621 Re_\tau^{11/7}.
\end{equation}
The relationship between the different Reynolds numbers and the corresponding velocity scales in a fully developed turbulent channel flow at relatively low Reynolds number, at which DNS are typically carried out, are summarized in table~\ref{Table:Re_relation2}. Comparison with the DNS described in \S\ref{sec:CpPI} where $Re_\tau=199.7$, $Re_b=3176$ and $\Repower=6500$ reveals the known little inaccuracy of the Deans' correlation at low values of $Re$.
\begin{table}
\caption{Relationship between different Reynolds numbers obtained by the Dean's formula~(\ref{eq:Cf_Reb}).}
\label{Table:Re_relation2}
\begin{center}
\begin{tabular}{l c c c c l}
\hline
$Re_\tau$ & $Re_b$ & $\Repower$ & $U^*_b / u^*_\tau$
& $\Upower^* / u^*_\tau$ & $\Upower^* / U^*_b$ \\
\hline \hline
100 & 1413 & 2171  & 14.4 & 21.7 & 1.54\\
150 & 2247 & 4105  & 15.0 & 27.4 & 1.83\\
200 & 3121 & 6451  & 15.6 & 32.3 & 2.07\\
300 & 4961 & 12199 & 16.5 & 40.7 & 2.46\\
450 & 7885 & 23070 & 17.5 & 51.3 & 2.93\\
600 &10954 & 36256 & 18.3 & 60.4 & 3.31\\
\hline
\end{tabular}
\end{center}
\end{table}

\begin{figure}
\centering
\psfrag{R}{$Re_b$}
\psfrag{Ub}{$U^*_{b,lam} / U^*_b$}
\includegraphics[width=0.7\textwidth]{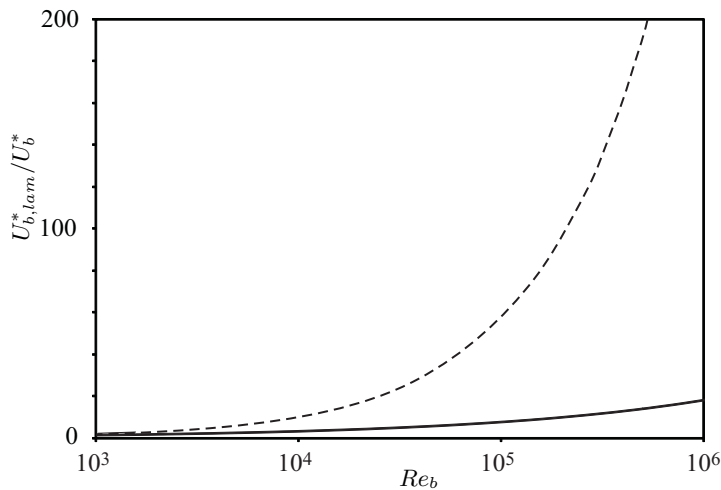}
\caption{The ratio of maximum achievable bulk flow velocity (i.e. the one in laminar flow) to the real turbulent bulk flow velocity as a function of bulk Reynolds number. Continuous line is the CPI case, Eq.~(\ref{eq:power_turb}), and dashed line is the CPG case, Eq.~(\ref{eq:Ub_CPG}).}
\label{fig:Rep-Reb}
\end{figure}

The last column of the table gives the ratio between the maximum achievable velocity $\Upower^*$ for the fixed power input (realized by the laminar flow state) and the actual turbulent bulk velocity $U^*_b$ according to Eq.~(\ref{eq:power_turb}).
This quantity is of particular interest in order to illustrate an important difference between CPI and CPG. The corresponding ratio between laminar and turbulent flow conditions at constant pressure gradient is given by:
\begin{equation}
\label{eq:Ub_CPG}
\frac{{U^*_{b,lam}}^{CPG}}{U^*_b} = 0.0102 Re_b^{3/4}.
\end{equation}
Both curves, corresponding to Eqns.~(\ref{eq:power_turb}) (where $U^*_{b,lam} = \Upower^*$) and (\ref{eq:Ub_CPG}), are plotted in Fig.~\ref{fig:Rep-Reb}. For CPG a drastic increase in flow rate is revealed if the flow is laminarized: for example, at the still relatively modest value of $Re_b \approx 80,000$, the flow rate is almost 50 times larger for laminar flow conditions under CPG. In contrast, at the same value of $Re$, CPI results in a potential flow rate increase of a factor seven only.  It should be noted that the laminar flow under CPG is driven by a power input that exceeds the one for the turbulent flow by factor 50. Therefore, the impressive effect of relaminarization found for CPG conditions cannot be expected in an engineering system if, as it is often the case, such system is limited by the available power. \cite{marusic-joseph-mahesh-2007} note that the CPG condition can be linked to an example given by Kolmogorov in which he addresses the increase in flow speed if the Volga river were laminar. The river -- in contrast to most technical applications -- is driven by an infinite reservoir of energy (if we assume an infinite reservoir of water that is feeding the river) such that the required increase in power input is not a limiting factor.

\end{document}